\documentclass[10pt,a4paper,onecolumn]{article}
\usepackage[T1]{fontenc}
\usepackage{marginnote}
\usepackage{graphicx}
\usepackage{xcolor}
\usepackage{authblk,etoolbox}
\usepackage{titlesec}
\usepackage{calc}
\usepackage{tikz}
\usepackage{hyperref}
\hypersetup{colorlinks,breaklinks,
            urlcolor=[rgb]{0.0, 0.5, 1.0},
            linkcolor=[rgb]{0.0, 0.5, 1.0}}
\usepackage{caption}
\usepackage{tcolorbox}
\usepackage{amssymb,amsmath}
\usepackage{ifxetex,ifluatex}
\usepackage{seqsplit}
\usepackage{xstring}

\usepackage{float}
\let\origfigure\figure
\let\endorigfigure\endfigure
\renewenvironment{figure}[1][2] {
    \expandafter\origfigure\expandafter[H]
} {
    \endorigfigure
}

\usepackage{fixltx2e} 
\usepackage[
  backend=biber,
]{biblatex}
\bibliography{paper.bib}

\let\textttOrig=\texttt
\def\texttt#1{\expandafter\textttOrig{\seqsplit{#1}}}
\renewcommand{\seqinsert}{\ifmmode
  \allowbreak
  \else\penalty6000\hspace{0pt plus 0.02em}\fi}

\makeatletter
\let\href@Orig=\href
\def\href@Urllike#1#2{\href@Orig{#1}{\begingroup
    \def\Url@String{#2}\Url@FormatString
    \endgroup}}
\def\href@Notdoi#1#2{\def\tempa{#1}\def\tempb{#2}%
  \ifx\tempa\tempb\relax\href@Urllike{#1}{#2}\else
  \href@Orig{#1}{#2}\fi}
\def\href#1#2{%
  \IfBeginWith{#1}{https://doi.org}%
  {\href@Urllike{#1}{#2}}{\href@Notdoi{#1}{#2}}}
\makeatother

\usepackage[top=3.5cm, bottom=3cm, right=1.5cm, left=1.5cm,
           headheight=2.2cm, reversemp, includemp, marginparwidth=4.0cm]{geometry}

\titleformat{\section}
  {\normalfont\sffamily\Large\bfseries}
  {}{0pt}{}
\titleformat{\subsection}
  {\normalfont\sffamily\large\bfseries}
  {}{0pt}{}
\titleformat{\subsubsection}
  {\normalfont\sffamily\bfseries}
  {}{0pt}{}
\titleformat*{\paragraph}
  {\sffamily\normalsize}

\usepackage{fancyhdr}
\makeatletter
\let\ps@plain\ps@fancy
\fancyheadoffset[L]{4.5cm}
\fancyfootoffset[L]{4.5cm}

\definecolor{linky}{rgb}{0.0, 0.5, 1.0}

\newtcolorbox{repobox}
   {colback=red, colframe=red!75!black,
     boxrule=0.5pt, arc=2pt, left=6pt, right=6pt, top=3pt, bottom=3pt}

\newcommand{\ExternalLink}{%
   \tikz[x=1.2ex, y=1.2ex, baseline=-0.05ex]{%
       \begin{scope}[x=1ex, y=1ex]
           \clip (-0.1,-0.1)
               --++ (-0, 1.2)
               --++ (0.6, 0)
               --++ (0, -0.6)
               --++ (0.6, 0)
               --++ (0, -1);
           \path[draw,
               line width = 0.5,
               rounded corners=0.5]
               (0,0) rectangle (1,1);
       \end{scope}
       \path[draw, line width = 0.5] (0.5, 0.5)
           -- (1, 1);
       \path[draw, line width = 0.5] (0.6, 1)
           -- (1, 1) -- (1, 0.6);
       }
   }

\patchcmd{\@maketitle}{center}{flushleft}{}{}
\patchcmd{\@maketitle}{center}{flushleft}{}{}
\patchcmd{\@maketitle}{\LARGE}{\LARGE\sffamily}{}{}
\def\maketitle{{%
  
  \AB@maketitle}}
\makeatletter
\renewcommand\AB@affilsepx{ \protect\Affilfont}
\renewcommand\AB@affilnote[1]{{\bfseries #1}\hspace{3pt}}
\renewcommand{\affil}[2][]%
   {\newaffiltrue\let\AB@blk@and\AB@pand
      \if\relax#1\relax\def\AB@note{\AB@thenote}\else\def\AB@note{#1}%
        \setcounter{Maxaffil}{0}\fi
        \begingroup
        \let\href=\href@Orig
        \let\texttt=\textttOrig
        \let\protect\@unexpandable@protect
        \def\thanks{\protect\thanks}\def\footnote{\protect\footnote}%
        \@temptokena=\expandafter{\AB@authors}%
        {\def\\{\protect\\\protect\Affilfont}\xdef\AB@temp{#2}}%
         \xdef\AB@authors{\the\@temptokena\AB@las\AB@au@str
         \protect\\[\affilsep]\protect\Affilfont\AB@temp}%
         \gdef\AB@las{}\gdef\AB@au@str{}%
        {\def\\{, \ignorespaces}\xdef\AB@temp{#2}}%
        \@temptokena=\expandafter{\AB@affillist}%
        \xdef\AB@affillist{\the\@temptokena \AB@affilsep
          \AB@affilnote{\AB@note}\protect\Affilfont\AB@temp}%
      \endgroup
       \let\AB@affilsep\AB@affilsepx
}
\makeatother

\renewcommand\Affilfont{\sffamily\small\mdseries}
\ifnum 0\ifxetex 1\fi\ifluatex 1\fi=0 
  \usepackage[T1]{fontenc}
  \usepackage[utf8]{inputenc}

\else 
  \ifxetex
    \usepackage{mathspec}
  \else
    \usepackage{fontspec}
  \fi
  \defaultfontfeatures{Ligatures=TeX,Scale=MatchLowercase}

\fi
\IfFileExists{upquote.sty}{\usepackage{upquote}}{}
\IfFileExists{microtype.sty}{%
\usepackage{microtype}
\UseMicrotypeSet[protrusion]{basicmath} 
}{}

\usepackage{hyperref}
\hypersetup{unicode=true,
            pdftitle={partycls: A Python package for structural clustering},
            pdfborder={0 0 0},
            breaklinks=true}
\usepackage{color}
\usepackage{fancyvrb}

\DefineVerbatimEnvironment{Highlighting}{Verbatim}{commandchars=\\\{\}}
\newenvironment{Shaded}{}{}

\newcommand{\ImportTok}[1]{#1}

\newcommand{\NormalTok}[1]{#1}
\newcommand{\OperatorTok}[1]{\textcolor[rgb]{0.40,0.40,0.40}{#1}}

\newcommand{\StringTok}[1]{\textcolor[rgb]{0.25,0.44,0.63}{#1}}

\let\addcontentslineOrig=\addcontentsline
\def\addcontentsline#1#2#3{\bgroup
  \let\texttt=\textttOrig\addcontentslineOrig{#1}{#2}{#3}\egroup}
\let\markbothOrig\markboth
\def\markboth#1#2{\bgroup
  \let\texttt=\textttOrig\markbothOrig{#1}{#2}\egroup}
\let\markrightOrig\markright
\def\markright#1{\bgroup
  \let\texttt=\textttOrig\markrightOrig{#1}\egroup}

\usepackage{graphicx,grffile}
\makeatletter
\def\maxwidth{\ifdim\Gin@nat@width>\linewidth\linewidth\else\Gin@nat@width\fi}
\def\maxheight{\ifdim\Gin@nat@height>\textheight\textheight\else\Gin@nat@height\fi}
\makeatother
\setkeys{Gin}{width=\maxwidth,height=\maxheight,keepaspectratio}
\IfFileExists{parskip.sty}{%
\usepackage{parskip}
}{
\setlength{\parindent}{0pt}
\setlength{\parskip}{6pt plus 2pt minus 1pt}
}
\ifx\paragraph\undefined\else
\let\oldparagraph\paragraph
\renewcommand{\paragraph}[1]{\oldparagraph{#1}\mbox{}}
\fi
\ifx\subparagraph\undefined\else
\let\oldsubparagraph\subparagraph
\renewcommand{\subparagraph}[1]{\oldsubparagraph{#1}\mbox{}}
\fi

\title{partycls: A Python package for structural clustering}

        \author[1]{Joris Paret}
          \author[2]{Daniele Coslovich}
    
      \affil[1]{Laboratoire Charles Coulomb (L2C), Université de Montpellier, CNRS,
        Montpellier, France $\quad\quad\quad$}
      
      \affil[2]{Dipartimento di Fisica, Università di Trieste, Italy}
  \date{\vspace{-5ex}}

\begin{document}
\maketitle

\marginpar{
  \sffamily\small

  {\bfseries DOI:} \href{https://doi.org/10.21105/joss.03723}{\color{linky}{10.21105/joss.03723}}

  \vspace{2mm}

  {\bfseries Software}
  \begin{itemize}
    \setlength\itemsep{0em}
    \item \href{https://github.com/openjournals/joss-reviews/issues/3723}{\color{linky}{Review}} \ExternalLink
    \item \href{https://github.com/jorisparet/partycls}{\color{linky}{Repository}} \ExternalLink
    \item \href{https://doi.org/10.5281/zenodo.5639407}{\color{linky}{Archive}} \ExternalLink
  \end{itemize}

  \vspace{4mm}
  \hrule
  \vspace{4mm}
  
  {\bfseries Editor:} \href{https://lucydot.github.io}{\color{linky}{Lucy Whalley}} \ExternalLink

  \vspace{2mm}
  
  {\bfseries Reviewers:} 
  \begin{itemize}
    \setlength\itemsep{0em}
    \item \href{https://github.com/govarguz}{\color{linky}{@govarguz}} \ExternalLink
    \item \href{https://github.com/FTurci}{\color{linky}{@FTurci}} \ExternalLink
  \end{itemize}
  
  \vspace{2mm}

  {\bfseries Submitted:} 27 August 2021\\
  {\bfseries Published:} 08 November 2021

  \vspace{2mm}
  {\bfseries License}\\
  Authors of papers retain copyright and release the work under a Creative Commons Attribution 4.0 International License (\href{https://creativecommons.org/licenses/by/4.0/}{\color{linky}{CC BY 4.0}}).
}

\hypertarget{summary}{%
\section{Summary}\label{summary}}

partycls is a Python framework for cluster analysis of systems of
interacting particles. By grouping particles that share similar
structural or dynamical properties, partycls enables rapid and
unsupervised exploration of the system's relevant features. It provides
descriptors suitable for applications in condensed matter physics and
integrates the necessary tools of unsupervised learning, such as
dimensionality reduction, into a streamlined workflow. Through a simple
and expressive interface, partycls allows one to open a trajectory file,
perform a clustering based on the selected structural descriptor, and
analyze and save the results with only a few lines of code.

\hypertarget{statement-of-need}{%
\section{Statement of need}\label{statement-of-need}}

Analysis of the local arrangements of atoms and molecules in dense
liquids and solids is crucial to understand their emergent physical
properties. This is particularly important in systems whose local
structure is heterogeneous, which include polycrystalline materials and
partially ordered systems, like semi-crystalline polymers (Ganda and
Stenzel 2020) or metastable liquids during crystal nucleation (Russo and
Tanaka 2016). Even more challenging is the case of glass-forming liquids
and glasses (Royall and Williams 2015), as these systems may display
locally favored structures, whose symmetry and chemical concentration
differ in a subtle way from the bulk.

Recently, unsupervised learning has emerged as a novel approach to
structural analysis of disordered materials (Reinhart et al. 2017;
Boattini, Dijkstra, and Filion 2019). In particular, clustering methods
based on simple observables, such as radial distribution functions, bond
angle distributions, and bond orientational parameters (BOP), provide
useful insight into the structural heterogenity of glassy systems
(Boattini et al. 2020; Paret, Jack, and Coslovich 2020). By grouping the
particles according to the similarity of their local structure, these
methods avoid the profileration of distinct structural signatures that
affects conventional methods, like Voronoi-based analysis (Tanemura et
al. 1977) or common neighbor analysis (CNA) (Honeycutt and Andersen
1987), as well as topological classification approaches (Malins et al.
2013; Lazar, Han, and Srolovitz 2015).

With the present code, we aim to provide a coherent numerical framework
for unsupervised learning of structural and dynamical features of
condensed matter systems. To the best of our knowledge, there is
currently no publicly available code that provides and integrates all
the necessary tools needed for this kind of analysis. Through a variety
of structural descriptors, dimensionality reduction methods, clustering
algorithms and filtering options, partycls makes it possible to discover
the key structural features of a system and to assess the robustness of
the results. It can readily analyze trajectories generated by molecular
dynamics simulation softwares, both within particle-based and
multi-scale approaches (see \emph{e.g.} Plimpton (1995), Anderson,
Glaser, and Glotzer (2020) and Guzman et al. (2019)). The code has
already been used in the context of a recent publication (Paret, Jack,
and Coslovich 2020) and can be easily extended. In particular, future
versions will implement descriptors that depend on both space \emph{and}
time, to learn about the dynamics of the system as well. Spatio-temporal
clustering may provide valuable insight into the heterogeneous dynamics
of amorphous materials (Berthier et al. 2011) and into particles'
trajectories in more extreme states of matter (Markidis et al. 2020).

\hypertarget{design}{%
\section{Design}\label{design}}

partycls is mostly written in Python, with a few parts coded in Fortran
90 for efficiency. It provides a simple and configurable workflow, from
reading the input trajectory, through the pre-processing steps, to the
final clustering results. It accepts several trajectory file formats, by
relying on optional third-party packages such as \texttt{MDTraj}
(McGibbon et al. 2015), which supports several well-known trajectory
formats, and \texttt{atooms} (Coslovich 2018), which makes it easy to
interface custom formats often used by in-house simulation codes. Thanks
to a flexible system of filters, it is possible to compute the
structural descriptors or perform the clustering on restricted subsets
of particles of the system, based on arbitrary particle properties. In
addition to its native descriptors, partycls also supports additional
structural descriptors via DScribe (Himanen et al. 2020).

\begin{figure}
\centering
\includegraphics{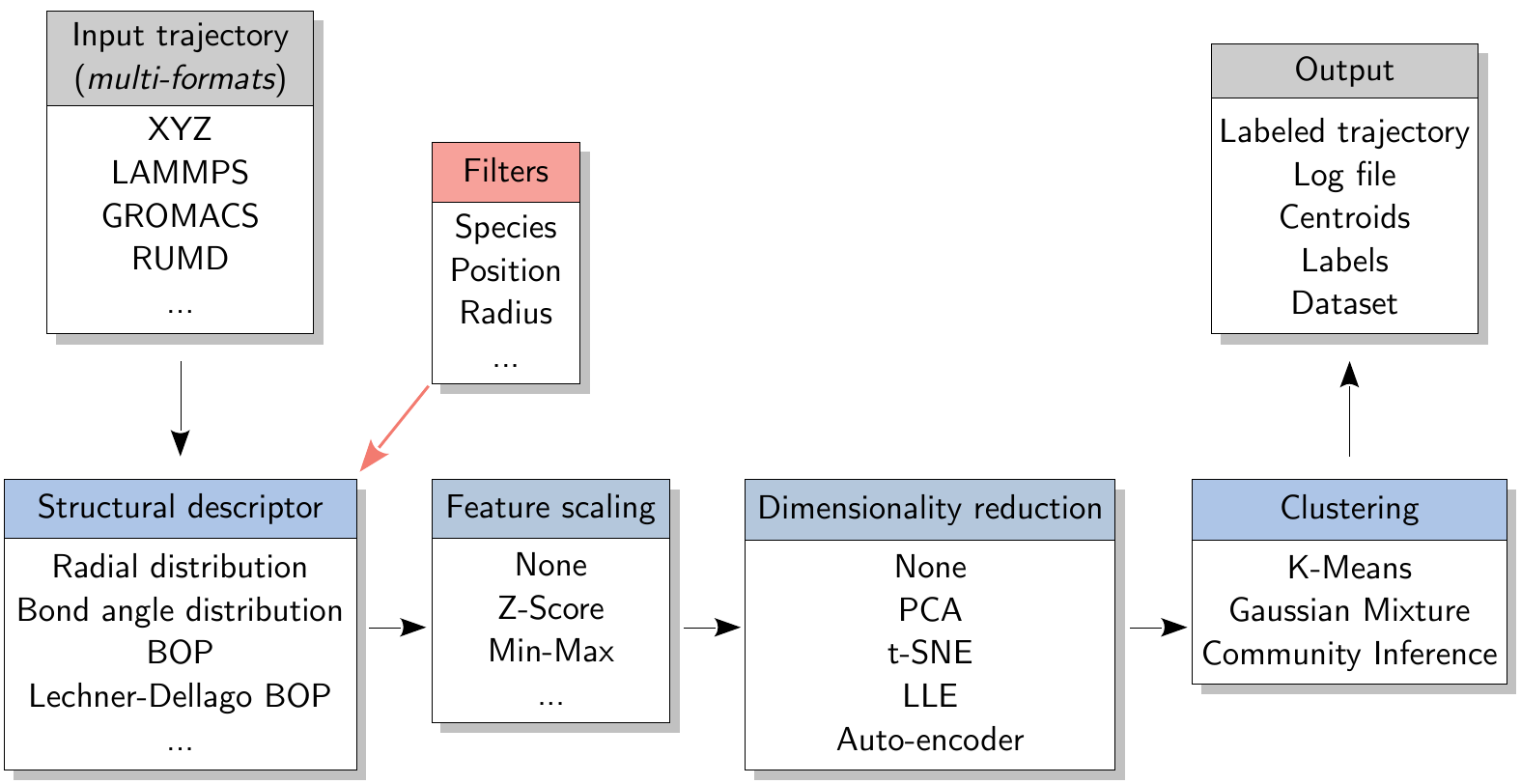}
\caption{The different steps to perform a structural clustering. The
input is a file written in any of the trajectory formats supported by
partycls. After selecting the structural descriptor and optional
filters, two key steps for pre-processing the data are possible: feature
scaling and dimensionality reduction. Finally, a clustering is performed
using the selected algorithm. Several output files are produced for
further analysis. \label{fig:workflow}}
\end{figure}

Some parts of the code act as a wrapper around functions of the machine
learning package \texttt{scikit-learn} (Pedregosa et al. 2011). This
allows non-experienced users to rely on the simplicity of partycls's
interface without any prior knowledge of this external package, while
experienced users can take full advantage of the many options provided
by \texttt{scikit-learn}. In addition, the code integrates the relevant
tools for distributional clustering, such as a community inference
method tailored to amorphous materials (Paret, Jack, and Coslovich
2020), and several helper functions, e.g.~for merging mixture models
(Baudry et al. 2010) and consistent centroid-based cluster labeling. A
simple diagram of the different steps and combinations to create a
custom workflow is shown in \autoref{fig:workflow}. A collection of
notebooks, with various examples and detailed instructions on how to run
the code, is available in the
\href{https://github.com/jorisparet/partycls}{partycls's repository}.

To maintain a consistent API as the code base evolves, partycls will
rigorously follow \href{https://semver.org/}{semantic versioning} as the
code is designed for maintainable extension. Future work will focus on
the underlooked case of dynamical clustering by implementing
time-dependent descriptors for individual particle trajectory.

\hypertarget{examples}{%
\section{Examples}\label{examples}}

As a simple example, we consider the detection of the grain boundaries
in a polycrystal formed by differently oriented FCC crystallites
(Rosenbrock et al. 2018). This is easily achieved even with a simple
radial descriptor, since the average radial distribution of particles at
the boundaries is different than the one of the crystal in the bulk. The
following short piece of code opens the input trajectory stored in the
file \texttt{grains.xyz}, computes the local radial distribution
functions of the particles, applies a standard Z-Score normalization on
the data, and finally performs a clustering using the Gaussian mixture
model (GMM) with \(K=2\) clusters (default):

\begin{Shaded}
\begin{Highlighting}[]
\ImportTok{from}\NormalTok{ partycls }\ImportTok{import}\NormalTok{ Workflow}

\NormalTok{wf }\OperatorTok{=}\NormalTok{ Workflow(}\StringTok{'grains.xyz'}\NormalTok{,}
\NormalTok{              descriptor}\OperatorTok{=}\StringTok{'gr'}\NormalTok{,}
\NormalTok{              scaling}\OperatorTok{=}\StringTok{'zscore'}\NormalTok{,}
\NormalTok{              clustering}\OperatorTok{=}\StringTok{'gmm'}\NormalTok{)}
\NormalTok{wf.run()}
\end{Highlighting}
\end{Shaded}

\begin{figure}
\centering
\includegraphics{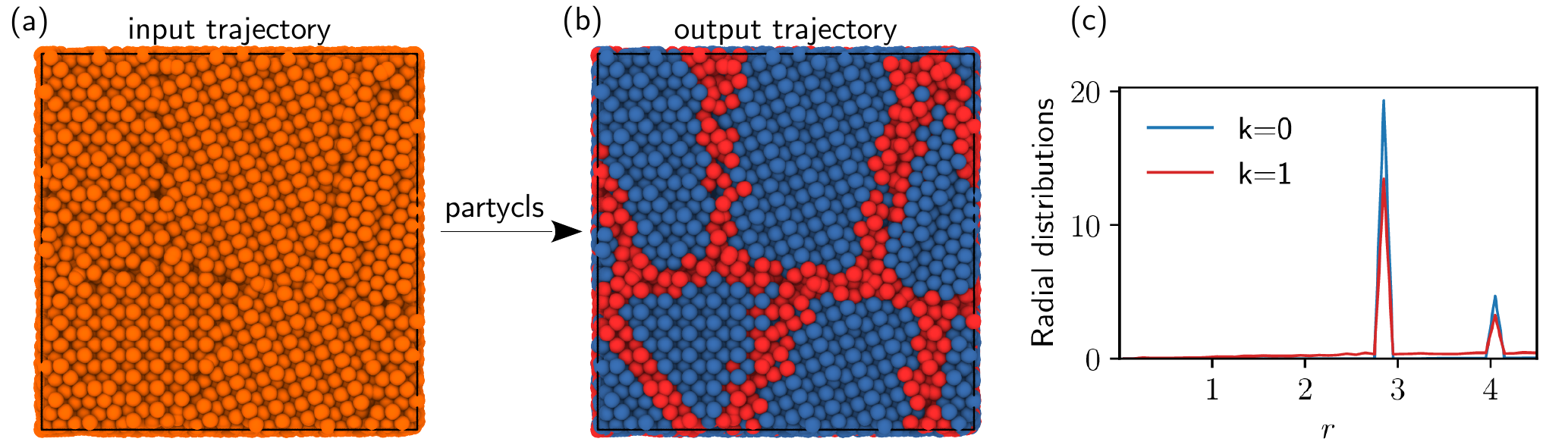}
\caption{(a) A polycrystalline material with differently oriented FCC
crystallites. (b) Using the individual radial distributions of the
particles as structural descriptor, the algorithm identifies the
crystalline domains (blue, \(k=0\)) and the grain boundaries (red,
\(k=1\)). (c) The radial distribution functions restricted to these two
clusters display a marked difference, with higher peaks for the
crystals. All 3D visualizations were performed with OVITO (Stukowski
2009). \label{fig:grains}}
\end{figure}

Each of these steps is easily tunable, so as to change the workflow with
little effort. The labels are available as a simple attribute of the
\texttt{Workflow} instance. Optionally, a set of output files can be
produced for further analysis, including a trajectory file with the
cluster labels. Quick visualization of the clusters, as in
\autoref{fig:grains}, is possible within partycls through optional
visualization backends.

The local structure of a glass-forming liquid provides a more
challenging bench-case, since the system is amorphous overall, but
subtle structural features emerge at low temperature. Here, we consider
a binary metallic alloy Cu\(_{64}\)Zr\(_{36}\), which shows a tendency
for local icosahedral arrangements around copper atoms (Soklaski et al.
2016). The fraction of atoms that form such locally favored structures
increases markedly when the system is cooled at low temperature. We use
LAMMPS (Plimpton 1995) to perform a molecular dynamics simulation using
an embedded atom potential. After a rapid quench from high temperature,
the supercooled liquid is annealed at \(T=900\)K. In the following piece
of code, we open a LAMMPS trajectory using \texttt{atooms} as backend,
we restrict the analysis to the copper atoms and use bond-angle
correlations and the K-Means algorithm to form the clusters:

\begin{Shaded}
\begin{Highlighting}[]
\ImportTok{from}\NormalTok{ partycls }\ImportTok{import}\NormalTok{ Trajectory, Workflow}
\ImportTok{from}\NormalTok{ partycls.descriptor }\ImportTok{import}\NormalTok{ BondAngleDescriptor}

\NormalTok{trajectory }\OperatorTok{=}\NormalTok{ Trajectory(}\StringTok{'cuzr_900K.dat'}\NormalTok{, fmt}\OperatorTok{=}\StringTok{'lammps'}\NormalTok{, backend}\OperatorTok{=}\StringTok{'atooms'}\NormalTok{)}
\NormalTok{descriptor }\OperatorTok{=}\NormalTok{ BondAngleDescriptor(trajectory)}
\NormalTok{descriptor.add_filter(}\StringTok{"species == 'Cu'"}\NormalTok{)}

\NormalTok{wf }\OperatorTok{=}\NormalTok{ Workflow(trajectory,}
\NormalTok{              descriptor}\OperatorTok{=}\NormalTok{descriptor,}
\NormalTok{              scaling}\OperatorTok{=}\StringTok{'zscore'}\NormalTok{,}
\NormalTok{              clustering}\OperatorTok{=}\StringTok{'kmeans'}\NormalTok{)}
\NormalTok{wf.run()}
\end{Highlighting}
\end{Shaded}

Here, we directly access classes for the trajectory and the structural
descriptor, and then pass them to the \texttt{Workflow} instance. Every
step of the workflow can also be performed manually by directly
instantiating the desired classes, without creating an instance of
\texttt{Workflow}.

\begin{figure}
\centering
\includegraphics{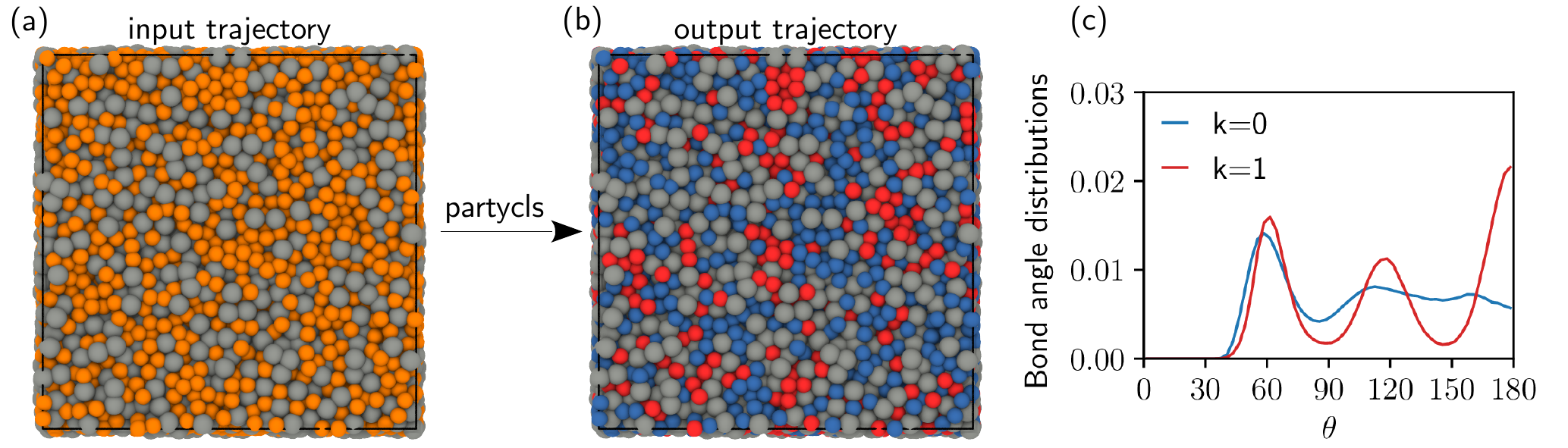}
\caption{(a) A glassy copper-zirconium alloy at \(T=900\)K. Copper and
zirconium atoms are colored in orange and grey, respectively. We focus
on the bond-angle distribution around the copper atoms only. (b) Copper
atoms are now colored blue (\(k=0\)) and red (\(k=1\)) based on their
cluster membership. Zirconium atoms (grey) are discarded from the
analysis. (c) Bond-angle distributions of the clusters.
\label{fig:cuzr}}
\end{figure}

In \autoref{fig:cuzr}, we see that the distribution of the cluster
\(k=1\) is similar to what is expected for icosahedral structural
environments, whereas that of the cluster \(k=0\) is flatter and thus
more disordered. This provides evidence of the local structural
heterogeneity of the system. Similar results have been obtained using
related clustering algorithms for simpler models of glass-forming
liquids based on Lennard-Jones interactions (Boattini et al. 2020;
Paret, Jack, and Coslovich 2020).

\hypertarget{acknowledgements}{%
\section{Acknowledgements}\label{acknowledgements}}

We thank Robert Jack for his contribution to the community inference
method, the merging of mixture models, and for his helpful comments.

\hypertarget{references}{%
\section*{References}\label{references}}
\addcontentsline{toc}{section}{References}

\hypertarget{refs}{}
\leavevmode\hypertarget{ref-anderson_hoomd_2020}{}%
Anderson, Joshua A., Jens Glaser, and Sharon C. Glotzer. 2020.
``HOOMD-Blue: A Python Package for High-Performance Molecular Dynamics
and Hard Particle Monte Carlo Simulations.'' \emph{Computational
Materials Science} 173: 109363.
\url{https://doi.org/10.1016/j.commatsci.2019.109363}.

\leavevmode\hypertarget{ref-baudry_combining_2010}{}%
Baudry, Jean-Patrick, Adrian E. Raftery, Gilles Celeux, Kenneth Lo, and
Raphaël Gottardo. 2010. ``Combining Mixture Components for Clustering.''
\emph{Journal of Computational and Graphical Statistics} 19 (2):
332--53. \url{https://doi.org/10.1198/jcgs.2010.08111}.

\leavevmode\hypertarget{ref-berthier_dynamical_2011}{}%
Berthier, Ludovic, Giulio Biroli, Jean-Philippe Bouchaud, Luca
Cipelletti, and Wim van Saarloos, eds. 2011. \emph{Dynamical
Heterogeneities in Glasses, Colloids, and Granular Media}. International
Series of Monographs on Physics. Oxford University Press.
\url{https://doi.org/10.1093/acprof:oso/9780199691470.001.0001}.

\leavevmode\hypertarget{ref-boattini_unsupervised_2019}{}%
Boattini, Emanuele, Marjolein Dijkstra, and Laura Filion. 2019.
``Unsupervised Learning for Local Structure Detection in Colloidal
Systems.'' \emph{The Journal of Chemical Physics} 151 (15): 154901.
\url{https://doi.org/10.1063/1.5118867}.

\leavevmode\hypertarget{ref-boattini_autonomously_2020}{}%
Boattini, Emanuele, Susana Marín-Aguilar, Saheli Mitra, Giuseppe Foffi,
Frank Smallenburg, and Laura Filion. 2020. ``Autonomously Revealing
Hidden Local Structures in Supercooled Liquids.'' \emph{Nature
Communications} 11 (1): 5479.
\url{https://doi.org/10.1038/s41467-020-19286-8}.

\leavevmode\hypertarget{ref-coslovich_daniele_2018_1183302}{}%
Coslovich, Daniele. 2018. \emph{atooms: A python framework for
simulations of interacting particles} (version 1.3.3). Zenodo.
\url{https://doi.org/10.5281/zenodo.1183302}.

\leavevmode\hypertarget{ref-Ganda_2020}{}%
Ganda, Sylvia, and Martina H. Stenzel. 2020. ``Concepts, Fabrication
Methods and Applications of Living Crystallization-Driven Self-Assembly
of Block Copolymers.'' \emph{Progress in Polymer Science} 101: 101195.
\url{https://doi.org/10.1016/j.progpolymsci.2019.101195}.

\leavevmode\hypertarget{ref-Guzman_Tretyakov_Kobayashi_Fogarty_Kreis_Krajniak_Junghans_Kremer_Stuehn_2019}{}%
Guzman, Horacio V., Nikita Tretyakov, Hideki Kobayashi, Aoife C.
Fogarty, Karsten Kreis, Jakub Krajniak, Christoph Junghans, Kurt Kremer,
and Torsten Stuehn. 2019. ``ESPResSo++ 2.0: Advanced Methods for
Multiscale Molecular Simulation.'' \emph{Computer Physics
Communications} 238 (May): 66--76.
\url{https://doi.org/10.1016/j.cpc.2018.12.017}.

\leavevmode\hypertarget{ref-dscribe}{}%
Himanen, Lauri, Marc O. J. Jäger, Eiaki V. Morooka, Filippo Federici
Canova, Yashasvi S. Ranawat, David Z. Gao, Patrick Rinke, and Adam S.
Foster. 2020. ``DScribe: Library of descriptors for machine learning in
materials science.'' \emph{Computer Physics Communications} 247: 106949.
\url{https://doi.org/10.1016/j.cpc.2019.106949}.

\leavevmode\hypertarget{ref-honeycutt_molecular_1987}{}%
Honeycutt, J. Dana., and Hans C. Andersen. 1987. ``Molecular Dynamics
Study of Melting and Freezing of Small Lennard-Jones Clusters.''
\emph{The Journal of Physical Chemistry} 91 (19): 4950--63.
\url{https://doi.org/10.1021/j100303a014}.

\leavevmode\hypertarget{ref-Lazar_Han_Srolovitz_2015}{}%
Lazar, Emanuel A., Jian Han, and David J. Srolovitz. 2015. ``A
Topological Framework for Local Structure Analysis in Condensed
Matter.'' \emph{Proceedings of the National Academy of Sciences} 112
(43): E5769--E5776. \url{https://doi.org/10.1073/pnas.1505788112}.

\leavevmode\hypertarget{ref-malins_identification_2013}{}%
Malins, Alex, Stephen R. Williams, Jens Eggers, and C. Patrick Royall.
2013. ``Identification of Structure in Condensed Matter with the
Topological Cluster Classification.'' \emph{The Journal of Chemical
Physics} 139 (23): 234506. \url{https://doi.org/10.1063/1.4832897}.

\leavevmode\hypertarget{ref-Markidis_Peng_Podobas_Jongsuebchoke_Bengtsson_Herman_2020}{}%
Markidis, Stefano, Ivy Peng, Artur Podobas, Itthinat Jongsuebchoke,
Gabriel Bengtsson, and Pawel Herman. 2020. ``Automatic Particle
Trajectory Classification in Plasma Simulations.'' In \emph{2020
Ieee/Acm Workshop on Machine Learning in High Performance Computing
Environments (Mlhpc) and Workshop on Artificial Intelligence and Machine
Learning for Scientific Applications (Ai4s)}, 64--71.
\url{https://doi.org/10.1109/MLHPCAI4S51975.2020.00014}.

\leavevmode\hypertarget{ref-McGibbon2015MDTraj}{}%
McGibbon, Robert T., Kyle A. Beauchamp, Matthew P. Harrigan, Christoph
Klein, Jason M. Swails, Carlos X. Hernández, Christian R. Schwantes,
Lee-Ping Wang, Thomas J. Lane, and Vijay S. Pande. 2015. ``MDTraj: A
Modern Open Library for the Analysis of Molecular Dynamics
Trajectories.'' \emph{Biophysical Journal} 109 (8): 1528--32.
\url{https://doi.org/10.1016/j.bpj.2015.08.015}.

\leavevmode\hypertarget{ref-paret_assessing_2020}{}%
Paret, Joris, Robert L. Jack, and Daniele Coslovich. 2020. ``Assessing
the Structural Heterogeneity of Supercooled Liquids Through Community
Inference.'' \emph{The Journal of Chemical Physics} 152 (14): 144502.
\url{https://doi.org/10.1063/5.0004732}.

\leavevmode\hypertarget{ref-scikit-learn}{}%
Pedregosa, F., G. Varoquaux, A. Gramfort, V. Michel, B. Thirion, O.
Grisel, M. Blondel, et al. 2011. ``Scikit-Learn: Machine Learning in
Python.'' \emph{Journal of Machine Learning Research} 12: 2825--30.

\leavevmode\hypertarget{ref-plimpton_fast_1995}{}%
Plimpton, Steve. 1995. ``Fast Parallel Algorithms for Short-Range
Molecular Dynamics.'' \emph{Journal of Computational Physics} 117 (1):
1--19. \url{https://doi.org/10.1006/jcph.1995.1039}.

\leavevmode\hypertarget{ref-reinhart_machine_2017}{}%
Reinhart, Wesley F., Andrew W. Long, Michael P. Howard, Andrew L.
Ferguson, and Athanassios Z. Panagiotopoulos. 2017. ``Machine Learning
for Autonomous Crystal Structure Identification.'' \emph{Soft Matter} 13
(27): 4733--45. \url{https://doi.org/10.1039/C7SM00957G}.

\leavevmode\hypertarget{ref-rosenbrock_structural_2018}{}%
Rosenbrock, Conrad W., Jonathan L. Priedeman, Gus L. W. Hart, and Eric
R. Homer. 2018. ``Structural Characterization of Grain Boundaries and
Machine Learning of Grain Boundary Energy and Mobility.''
\emph{arXiv:1808.05292 {[}Cond-Mat, Physics:Physics{]}}, August.
\url{http://arxiv.org/abs/1808.05292}.

\leavevmode\hypertarget{ref-Royall_Williams_2015}{}%
Royall, C. Patrick, and Stephen R. Williams. 2015. ``The Role of Local
Structure in Dynamical Arrest.'' \emph{Physics Reports} 560 (February):
1--75. \url{https://doi.org/10.1016/j.physrep.2014.11.004}.

\leavevmode\hypertarget{ref-Russo_Tanaka_2016}{}%
Russo, John, and Hajime Tanaka. 2016. ``Crystal Nucleation as the
Ordering of Multiple Order Parameters.'' \emph{The Journal of Chemical
Physics} 145 (21): 211801. \url{https://doi.org/10.1063/1.4962166}.

\leavevmode\hypertarget{ref-soklaski_locally_2016}{}%
Soklaski, Ryan, Vy Tran, Zohar Nussinov, K. F. Kelton, and Li Yang.
2016. ``A Locally Preferred Structure Characterises All Dynamical
Regimes of a Supercooled Liquid.'' \emph{Philosophical Magazine} 96
(12): 1212--27. \url{https://doi.org/10.1080/14786435.2016.1158427}.

\leavevmode\hypertarget{ref-ovito}{}%
Stukowski, Alexander. 2009. ``Visualization and Analysis of Atomistic
Simulation Data with OVITOthe Open Visualization Tool.'' \emph{Modelling
and Simulation in Materials Science and Engineering} 18 (1): 015012.
\url{https://doi.org/10.1088/0965-0393/18/1/015012}.

\leavevmode\hypertarget{ref-tanemura_geometrical_1977}{}%
Tanemura, Masaharu, Yasuaki Hiwatari, Hirotsugu Matsuda, Tohru Ogawa,
Naofumi Ogita, and Akira Ueda. 1977. ``Geometrical Analysis of
Crystallization of the Soft-Core Model*).'' \emph{Progress of
Theoretical Physics} 58 (4): 1079--95.
\url{https://doi.org/10.1143/PTP.58.1079}.

\end{document}